\documentclass[twocolumn,floatfix,showpacs,prl,aps]{revtex4}
\usepackage[applemac]{inputenc}
\usepackage[T1]{fontenc} 
\pdfoutput=1
\usepackage{amssymb}
\usepackage{enumerate}
\usepackage{braket}
 \usepackage{amsmath} 
\usepackage{subfigure}
\usepackage{lipsum}
\usepackage{amsfonts} 
\usepackage{amssymb, mathrsfs}
\usepackage{amsmath}
\usepackage{bm}
\usepackage[pdftex]{graphicx}
\usepackage{footnote}

\def\beq{\begin{equation}}
\def\eeq{\end{equation}}
\def\bsp{\begin{split}}
\def\esp{\end{split}}
\def\bea{\begin{eqnarray}}
\def\eea{\end{eqnarray}}
\def\ba{\begin{array}}
\def\ea{\end{array}}

\def\l.{\left.}
\def\r.{\right.}

\def\part{\partial}

\begin{document}

\preprint{UdeM-GPP-TH-14-235}
\preprint{arXiv:1407.1457}
\title{Negative mass bubbles in de Sitter space-time.}
\author{Saoussen Mbarek\footnote{present address: Department of Physics, University of Waterloo}}
\email{smbarek@uwaterloo.ca}
\author{M. B. Paranjape} 
\email{paranj@lps.umontreal.ca}
\affiliation{Groupe de physique des particules, D\'epartement de physique,
Universit\'e de Montr\'eal,
C.P. 6128, succ. centre-ville, Montr\'eal, 
Qu\'ebec, Canada, H3C 3J7 }

\begin{abstract}

\section{Abstract}  
We study the possibility of the existence of negative mass bubbles within a de Sitter space-time background with matter content corresponding to a perfect fluid.  It is shown that there exist configurations of the perfect fluid, that everywhere satisfy  the dominant energy condition, the Einstein equations and the equations of hydrostatic equilibrium, however asymptotically approach the exact solution of Schwarzschild-de Sitter space-time with a negative mass.
\end{abstract}


\maketitle


{\it{Introduction}-}
The Schwarzschild solution \cite{s} is an exact vacuum solution of the Einstein equations which contains one free parameter, the mass.  The solution is singular at Schwarzschild coordinate $r=0$.  For a positive mass, the solution is called a black hole, and the singularity is hidden behind a null surface called the event horizon.  The singularity, rather than being concentrated at a point,  actually occurs on a future space-like hypersurface.   If an observer stays outside the event horizon, he is safe from the singularity.  However, if he crosses over the null surface and enters the black hole he cannot avoid the singularity, just as he cannot avoid aging, and in finite time he is ripped apart by the infinite gravitational stresses that occur at the singularity.  On the other hand, for a negative mass, the singularity  is naked as it has no event horizon cloaking it.  However, it is point like, and occurs at a fixed spatial position.  An observer can view the singularity from a distance,  and in principle, every observer who chooses not to impinge on the singularity can avoid it forever.  Indeed, the singularity of the negative mass  Schwarzschild solution appears to be relatively benign compared to the one of a positive mass.

We are commonly exposed to singularities of the type exhibited by the negative mass Schwarzschild geometry.  For example consider a point charge in electrodynamics.  The corresponding electric field is in principle, singular at the location of the point charge, and the total energy is infinite.  We are content with the understanding that from a large distance, a singular point charge is harmless, we can simply avoid it, and at close distance we expect that the singular nature of the charge will be smoothed out by a concentrated but non-singular charge density.  The analogous situation in the context of general relativity and the negative mass Schwarzschild solution, is to ask if there is a non-singular distribution of energy-momentum density that smooths out its singularity.  This question was analyzed in \cite{bp} and it was concluded that for asymptotically Minkowski space-time, it is actually easy to remove the singularity in the metric, but only at the expense of introducing other pathologies in the system, that are untenable. 

The physical definition of the mass is subjective.  It can mean the total ADM mass \cite{adm} or the Bondi mass  \cite{Bondi} or the Komar mass \cite{Komar}, depending on the symmetries and asymptotics of the metrics involved.  We take the very physical view point of defining the mass as that parameter which would appear in the Newton dynamical equations in the limit of slowly moving sources and weak gravitational forces.  This mass is exactly the Schwarzschild-like mass parameter that appears in the metric, as shown by Bondi in \cite{Bondi2} and Luttinger \cite {Luttinger}.

A non-singular matter distribution that is  physically reasonable and  not pathological, must correspond to an energy momentum tensor that is non singular but additionally,  satisfies everywhere the dominant energy condition \cite{dec}.   This condition is equivalent to the statement that no Lorentz observer can observe the energy-momentum to be moving out of the future directed light cone.    It was observed in \cite{bp}, that any non-singular matter distribution, that asymptotically approaches the negative mass Schwarzschild metric, must violate the dominant energy condition.  The reason is because of the positive energy theorem  \cite{pet}, which states that in an asymptotically Minkowski space-time, any energy momentum tensor that everywhere satisfies the dominant energy condition must give rise to a metric that has positive ADM mass \cite{adm}.  The negative mass Schwarzschild metric evidently has a negative ADM mass.  Therefore, there  cannot  exist a smooth configuration of energy-momentum that everywhere satisfies the dominant energy condition and asymptotically approaches the negative mass Schwarzschild metric.  Thus the negative mass Schwarzschild solution remains an unphysical solution and the question of its meaning is still unanswered.

However, it is possible to evade the positive energy theorem by relaxing any of the assumptions that it requires.  In \cite{bp}, it was pointed out, that removing the condition that the space-time be asymptotically Minkowski is a useful reduction.  In this letter we consider asymptotically de Sitter space times.  Positive energy theorems and the analysis of the positivity of the mass have been done in such space-times \cite{kt,sh,ad,nam,cjk}, but our analysis avoids their conclusions.  The papers \cite{kt}  prove the positivity of the mass associated with the conformal time translation Killing vector.  Although positive, this mass is time dependent as far as the static time coordinate is concerned, and thus not what we would like to consider as the mass.  The paper \cite{sh} and \cite{nam} prove the positivity of the AD mass as defined by \cite{ad}, but only if it is assumed to be positive on an initial surface.  We do not make that assumption.  The paper  \cite{cjk}  makes stringent conditions on the asymptotic space-time which also guarantees the positivity of the mass, however we do not impose the same conditions.  Finally the paper \cite{le} gives a nice counterexample to the conjecture that locally asymptotically flat space-times must also have positive mass.  But none of these papers consider our specific, physical problem of a perfect fluid in an asymptotically de Sitter space-time.  

The Schwarzschild-de Sitter solutions are  exact solutions of the Einstein equations in the presence of a constant, background energy-momentum density (or equivalently with cosmological constant), which correspond to an exponentially evolving space-time geometry with a singular, point like mass.  Explicitly the metric is given by
\bea
ds^2&=&\left(1-\frac{2(\Lambda r^3/6+M)}{r}\right)dt^2\nonumber\\&-&\left(1-\frac{2(\Lambda r^3/6+M)}{r}\right)^{-1}dr^2-r^2d\Omega^2\label{sds}.
\eea

 The solutions contain two parameters, the value of the mass $M$, and the value of the constant energy density (or cosmological constant), $\Lambda$.  Either can be positive or negative.  Negative $\Lambda$  gives rise to anti-de Sitter space-time, which, although of great interest recently \cite{m}, already on its own does not satisfy the dominant energy condition.   Positive  $\Lambda$ gives rise to de Sitter space-time.  The mass parameter, $M$, as expected describes a positive or negative point-like mass depending on its sign.  The positive mass is hidden behind an event horizon, while the negative mass singularity is naked, and occurs at a fixed spatial point.   The positive energy theorem \cite{pet} is not applicable since the space-time is not asymptotically Minkowski.   The negative mass singularity can be smoothed out in a de Sitter background while maintaining the dominant energy condition everywhere, as was shown in \cite{bp}, however, no attempt was made to find any type of energy-momentum which could give rise to the deformation that is required.   In this paper we  show that energy-momentum corresponding to a perfect fluid can be used to provide a suitable deformation.  We find bubble like configurations which are non-singular, satisfy the dominant energy condition everywhere and give rise to a metric that asymptotically approaches the negative mass Schwarzschild-de Sitter geometry.  

{\it{Equations of motion}-}
The metric of a spherically symmetric space-time can be taken as
\beq
ds^2=B(r)dt^2-A(r)dr^2-r^2d\theta ^2-r^2{sin^2\theta} d\phi ^2\label{1}.
\eeq 
This metric is required to be a solution of the Einstein field equations:
\beq
G_{\mu \nu}=8\pi T_{\mu \nu}\label{2}
\eeq 
where we take that $c=G=1$. It has been commonly understood that the left hand side of the equation Eqn.\eqref{2} depends on the geometry of the space-time and the right hand side depends on the matter content of the universe and its distribution.  
We wish to solve the Einstein field equations for matter corresponding to a perfect fluid, where the metric asymptotically approaches the exact solution corresponding to a {\it negative} mass Schwarzschild-de Sitter geometry.  A perfect fluid is described by a stress energy tensor of the form
\beq
T_{\mu \nu}=-p g_{\mu \nu}+(p+\rho) U_\mu U_\nu\label{3}
\eeq 
where $p$ and $\rho$ are respectively the pressure and the density of the perfect fluid when $U_\mu$ is its four vector velocity. We choose a frame of reference where the fluid is at rest, then this four vector velocity is defined by $U_\mu=(\sqrt{B},0,0,0)$,  so that it obeys  $U_\mu U^\mu=1$.
This means that the only non-zero terms of the stress-energy tensor are:
 \beq
T_{00}=B\, \rho 
\quad
T_{11}=A\, p 
\quad
T_{22}=A\, p\, r^2 
\quad
T_{33}=\sin^2\theta \, T_{22} \label{7}
\eeq 
The Einstein equations in the presence of a perfect fluid are well known, for example in \cite{w}, or in any other standard reference on general relativity, however, we record them here for completeness.  We get three independent  equations:
\beq
\frac{B''}{2A}-\frac{B'}{4A}\left[\frac{B'}{B}+\frac{A'}{A}\right]+\frac{B'}{rA}=4\pi(3p+\rho)B \label{14}
\eeq 
\beq
-\frac{B''}{2B}+\frac{B'}{4B}\left[\frac{B'}{B}+\frac{A'}{A}\right]+\frac{A'}{rA}=4\pi(\rho-p)A \label{15}
\eeq 
\beq
1-\frac{r}{2A}\left[\frac{B'}{B}-\frac{A'}{A}\right]-\frac{1}{A}=4\pi(\rho-p)r^2 \label{16}
\eeq 
These  equations are underdetermined as they correspond to three equations among four fields in one independent variable, however, we will  use them to determine the pressure $p$, the density $\rho$, the time component of the metric $B$ as a function of the gravitational effective mass function $m(r)$.    A fourth equation corresponding to the condition of hydrostatic equilibrium is actually not an  independent equation, and in fact a consequence of the covariant conservation of the energy-momentum tensor, which in turn is equivalent to the consistency requirement of the Einstein equation with the Bianchi identities.  It corresponds to:
\beq
\frac{B'}{B}=\frac{-2p}{p+\rho} \label{17}.
\eeq 
Usually, to render the system of equations deterministic, a further equation is introduced corresponding to the equation of state, $p=p(\rho )$, which then picks out a unique solution.  However we will not impose such an equation, we will instead parametrize 
\beq
A=\left(1-\frac{2m(r)}{r}\right)^{-1},\label{A}
\eeq 
and choose $m(r)$ explicitly, which has the interpretation as the mass inside the radius $r$.  We then find the solution (numerically) for $p(r)$, $\rho(r)$ and $B(r)$ dependent on our choice of $m(r)$.  It will be clear from our solution that an equation of state connecting $p(r)$ and $\rho(r)$ must exist implicitly.

We can calculate the expression for the density through the operation $\frac{\rm Eqn.\eqref{14}}{2B}+\frac{\rm Eqn.\eqref{15}}{2A}+\frac{\rm Eqn.\eqref{16}}{r^2}$ to obtain
\beq
\rho(r)=\frac{1}{4\pi}\left(\frac{m'(r)}{r^2}\right) \label{18}
\eeq 
and we can also get an equation for the time component of the metric $B$ using the operation $\frac{\rm Eqn.\eqref{14}}{2B}+\frac{\rm Eqn.\eqref{15}}{2A}-\frac{\rm Eqn.\eqref{16}}{r^2}$ to obtain
\beq
\frac{B'}{B}=\frac{2}{r^2}\left[m(r)+4\pi r^3p\right]A .\label{19}
\eeq 
Finally, we can combine Eqn.\eqref{19} with Eqns.\ \eqref{14},\eqref{15},\eqref{16} to get
\bea
p'(r)=\frac{4\pi r^2}{2m(r)-r}p^2(r)+\frac{m(r)+rm'(r)}{r\left(2m(r)-r\right)}p(r)+\nonumber\\
+\frac{m'(r)m(r)}{4\pi r^3\left(2m(r)-r\right)} .\label{20}
\eea

To be able to numerically solve the equations, we need to define the gravitational effective mass function $m(r)$.  We obtain a smooth function through the following steps. We start with the function
 \begin{align}
\tilde m(r)=
  \begin{cases} 
  0      & \text{if }\ r < x \\
  a(r-x)^3      & \text{if }\ x < r < y \\
  \frac{\Lambda r^3}{6}-M     & \text{if }\ r > y\\
  \end{cases}\label{21}
 \end{align}
 where $x$, $y$ and $a$ are parameters which are constrained by imposing that the interpolation be ${\cal C}^2$ at $x$ and $y$, and we have redefined $M\rightarrow -M$, see \footnote{We have replaced $M$ in Eqn.\eqref{sds} with $-M$ here and from now on, as we are interested in a negative mass configuration.}.  This implies
\beq
  y=\sqrt{\frac{6M}{\Lambda x}} \quad\quad a=\frac{\Lambda y^2}{6(y-x)^2}      
 \label{22}.
 \eeq
 In the region $r>y$ the space-time becomes the exact negative mass Schwarzschild-de Sitter space-time. Then we define a  ${\cal C}^\infty$ interpolation by
 \bea
 m(r)&=\left(\frac{1+\tanh\omega(r-x)}{2}\right)  \left(\frac{1-\tanh\omega(r-y)}{2}\right)a(r-x)^3\nonumber\\     & + \left(\frac{1+\tanh\omega(r-y)}{2}\right) \left(\frac{\Lambda r^3}{6}-M\right) \label{m}
 \eea
 for a suitable value of $\omega$.   
 
 Our choice for $\tilde m(r)$ is motivated by the analysis done in \cite{bp} where it was shown that the dominant energy condition is satisfied if:
\begin{equation}
\frac{d}{dr}\left(\frac{\tilde m'(r)}{r^2} \right) \leq 0 \quad\quad \frac{d}{dr}\left(\tilde m'(r)r^2 \right) \geq 0 \label{EC2}
\end{equation}
Although the analysis done in \cite{bp} does not exactly apply  since there it was further assumed that  $B=1-2 \tilde m(r)/r$ which we do not assume here, $\tilde m(r)$  satisfies the conditions of Eqn.\eqref{EC2}.  The dominant energy condition for energy-momentum described by an ideal fluid simply requires
\beq
\rho(r)\ge 0 \quad\quad \rho(r)\ge |p(r)|\label{dec} 
\eeq
which we will find are respected by our solution of the field equations.
\begin{figure}[ht]
\centerline{\includegraphics[width=0.9\linewidth]{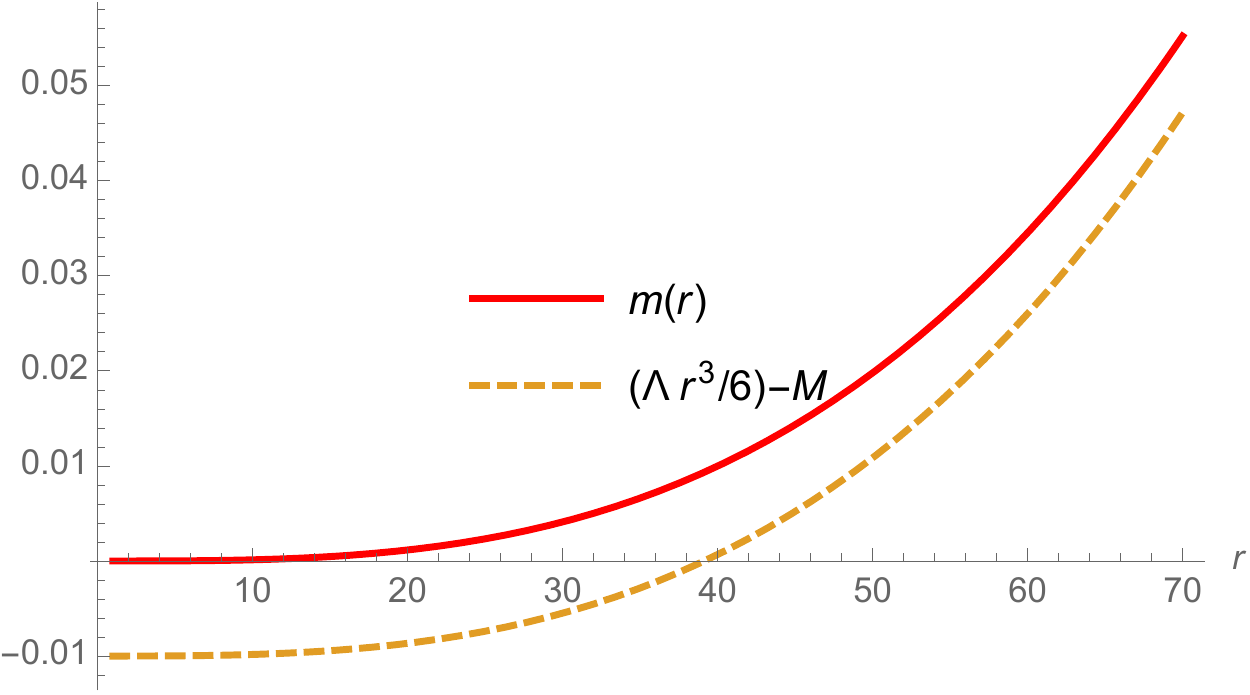}} \caption{\label{fig0} (color online) $m(r)$ for the parameters $x=.1$, $\Lambda=.001$, $M=.01$, $y\approx245$.}
\end{figure}

The de Sitter horizon is given by
\beq
2m(r)-r=0\label{dh}
\eeq
which must be at a much larger radius than the size of the bubble given by $y$ for a sensible configuration.  The solution of Eqn.\eqref{dh} must occur when $r$ is in the Schwarzschild-de Sitter region of the space-time, $r\gg y$.  Neglecting $M$ then gives
\beq
2( \frac{\Lambda r^3}{6}-M )-r=0\Rightarrow r\approx\sqrt{\frac{3}{\Lambda}}.
\eeq
Comparing this with the expression for $y$ from Eqn.\eqref{22} yields
\beq
\sqrt{\frac{6M}{\Lambda x}} \ll \sqrt{\frac{3}{\Lambda}}.
\eeq

\begin{figure}[ht]
\centerline{\includegraphics[width=0.9\linewidth]{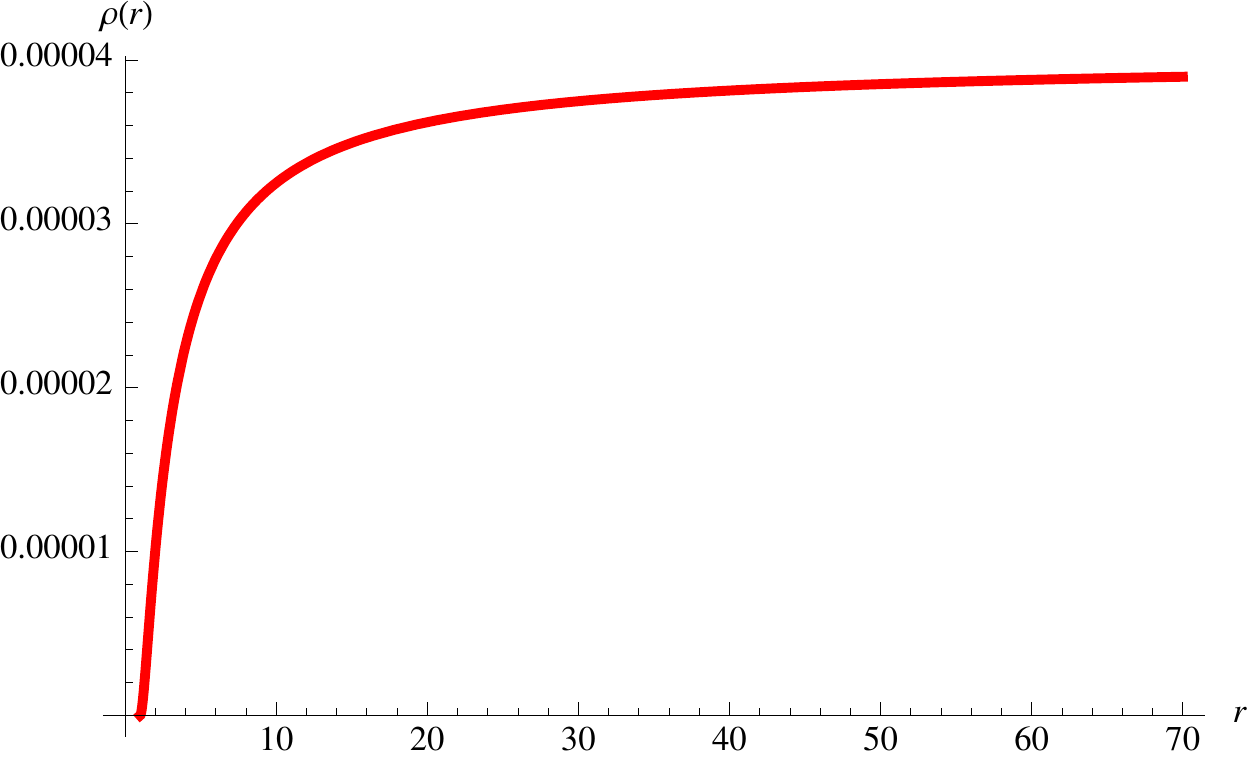}} \caption{\label{fig2} (color online) Density $\rho(r)$ for the parameters $x=.1$ $G=1$, $\Lambda=.001$, $M=.01$, $y\approx245$.}
\end{figure}

\begin{figure}[ht]
\centerline{\includegraphics[width=0.9\linewidth]{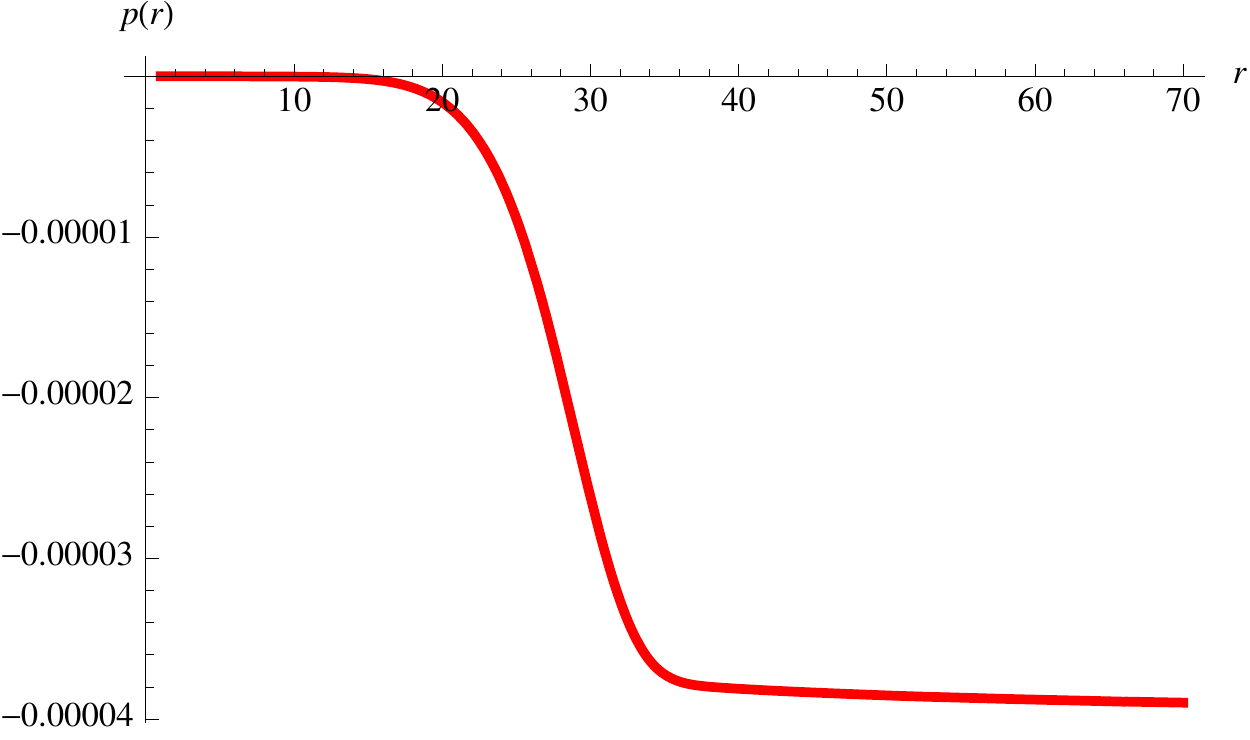}} \caption{\label{fig1} (color online) Pressure $p(r)$ for the parameters $x=.1$ $G=1$, $\Lambda=.001$, $M=.01$, $y\approx245$.}\label{p}
\end{figure}

Using these conditions combined with the expression of $m(r)$, we get the numerical solution for the density and the pressure of the perfect fluid, $\rho(r)$ and $p(r)$, given respectively in Fig.(\ref{fig2}) and Fig.(\ref{fig1}).


\begin{figure}[ht]
\centerline{\includegraphics[width=0.9\linewidth]{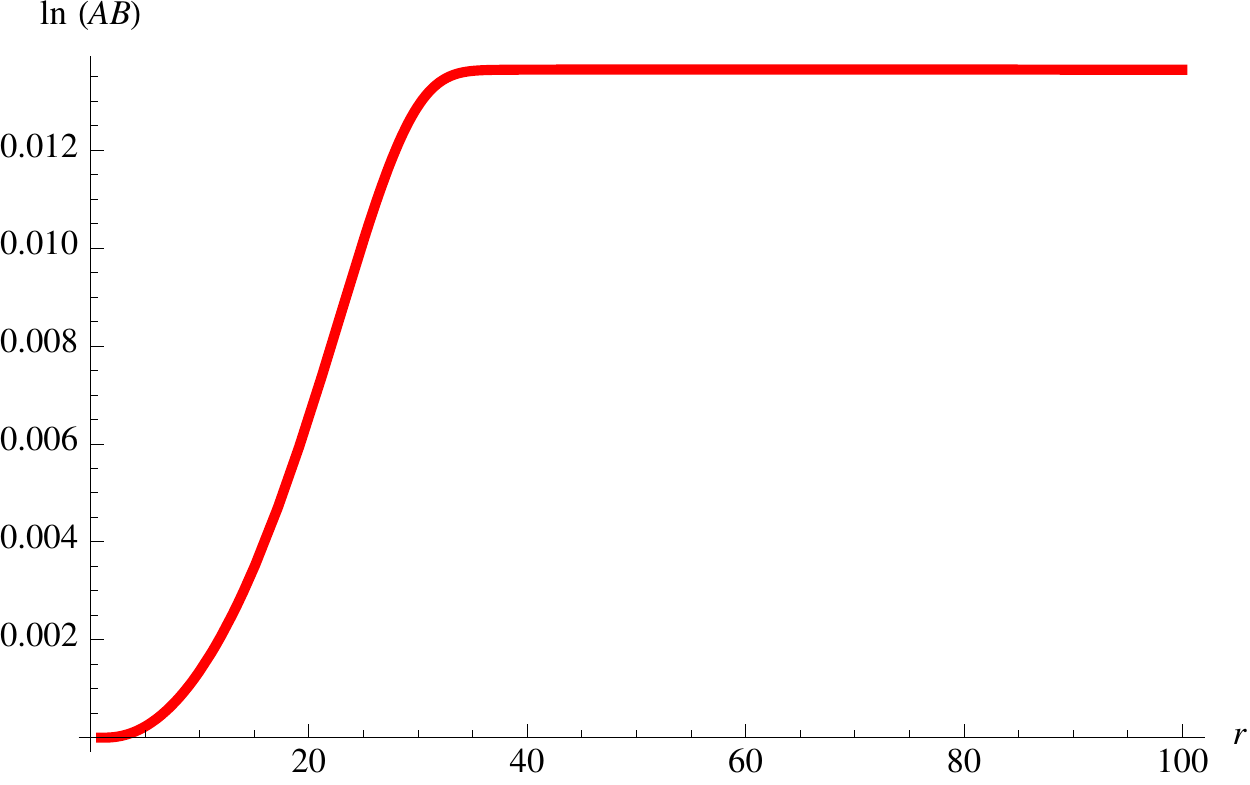}} \caption{\label{fig3} (color online) $\ln AB$ for the parameters $x=.1$ $G=1$, $\Lambda=.001$, $M=.01$, $y\approx245$.}
\end{figure}

We note that the pressure is negative and interpolates from $p(r)=0$ in the interior where the space-time is Minkowski to $p(r)=-\rho(r)$, which is the normal solution for de Sitter space-time.  The numerical solution for $B(r)$ is not particularly simple to obtain because of numerical instability.  Instead of directly numerically integrating Eqn.\eqref{17}, or Eqn.\eqref{19}, it is better to find an expression for $AB$ , which comes straitforwardly from adding $\frac{\rm Eqn.\eqref{14}}{A}+\frac{\rm Eqn.\eqref{15}}{B}$:
\beq
\left(\ln{AB}\right)'=8\pi G\left(\rho(r) +p(r)\right)r A.\label{lnAB}
\eeq
Numerically integrating Eqn.\eqref{lnAB} yields the curve in Fig.(\ref{fig3}).

Thus our solution for the metric is given by Eqn.\eqref{A} for $A(r)$, where $m(r)$ is explicitly chosen as in Eqn.\eqref{m} and $B(r)$ obtained from numerically integrating Eqn.\eqref{lnAB} for $\ln(AB)$ and then trivially solving  $B(r)=e^{\ln(AB)}/A$.  To perform this integration, we need to first find $\rho(r)$, which is given by Eqn.\eqref{18} and $p(r)$, which is obtained by numerically integrating Eqn.\eqref{20} and shown in Fig.\eqref{p}.  
  
From Fig.(\ref{fig1}) and Fig.(\ref{fig2}) we can see the combination $\left(\rho(r) +p(r)\right)$ vanishes as $r$ becomes large.   Thus the expectation is that $AB$ becomes a constant for large $r$ as can be seen from Fig.(\ref{fig3}).  This constant can be taken to be 1, although it is not so in Fig.(\ref{fig3}) as the integration of Eqn.\eqref{lnAB} is only determined up to a constant.  Notice that the Einstein equations, Eqns.\eqref{14},\eqref{15},\eqref{16},\eqref{17} all are homogeneous in $B$.  If $B$ is a solution then so is $\alpha B$ for any constant (positive) $\alpha$.  Thus asymptotically, $B=1/A$ and the solution approaches the exact negative mass Schwarzschild-de Sitter solution. 

\begin{figure}[ht]
\centerline{\includegraphics[width=0.9\linewidth]{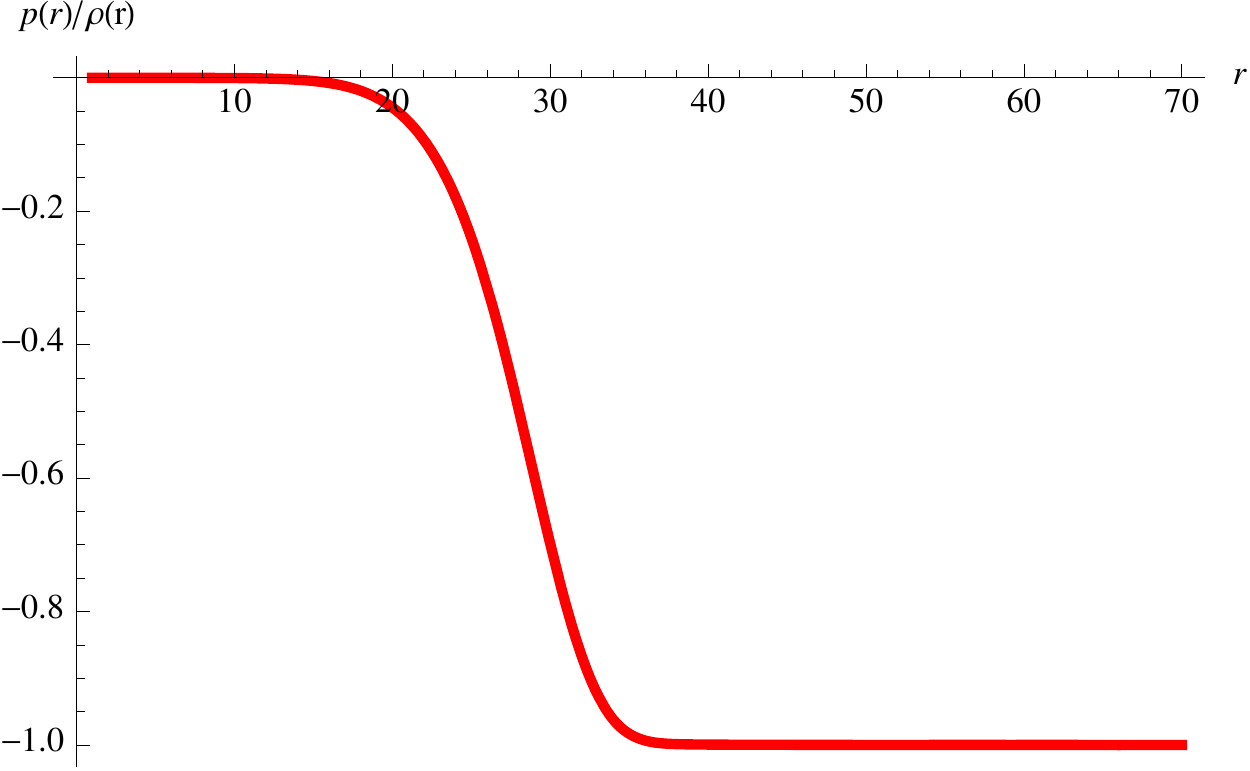}} \caption{\label{fig4} (color online) $p(r)/\rho(r)$ for the parameters $x=.1$ $G=1$, $\Lambda=.001$, $M=.01$, $y\approx245$.}
\end{figure}

If we plot $p(r)/\rho(r)$ we find the curve given by Fig.(\ref{fig4}), which is bounded between 0 and -1.  It is clear then, within the limits of numerical accuracy, that the constraints of the dominant energy condition, Eqn.\eqref{dec} are satisfied. 

{\it Conclusions-}  We have shown that there exist very physical configurations of an ideal fluid which give rise to solutions of the Einstein equations that correspond asymptotically to {\it negative} mass Schwarzschild-de Sitter space times.  The energy-momentum tensor that gives rise to such space times is perfectly physical, it everywhere satisfies the dominant energy condition.  Since the space time is not asymptotically flat, we evade the positive energy theorems which would not allow for negative mass.  Negative mass configurations therefore can exist in de Sitter backgrounds, exactly as have been proposed for the inflationary phase of the early universe.  If a mechanism for production of pairs of particles with positive and negative mass can be determined, in the early universe there would be a plasma of positive and negative mass particles.  Such a plasma would in principle cause an effective screening of  gravitational waves, being essentially opaque for frequencies below the plasma frequency.

{\it  Acknowledgments-} We thank NSERC of Canada and MUTAN of Tunisia for financial support. 


\end{document}